# Advanced Neural Network Architecture for Enhanced Multi-Lead ECG Arrhythmia Detection through Optimized Feature Extraction

Bhavith Chandra Challagundla[1]

[1]*Student, Computational Intelligence, School of Computing, SRMIST*

## 1. Abstract

Cardiovascular diseases are a pervasive global health concern, contributing significantly to morbidity and mortality rates worldwide. Among these conditions, arrhythmia, characterized by irregular heart rhythms, presents formidable diagnostic challenges. This study introduces an innovative approach utilizing deep learning techniques, specifically Convolutional Neural Networks (CNNs), to address the complexities of arrhythmia classification. Leveraging multi-lead Electrocardiogram (ECG) data, our CNN model, comprising six layers with a residual block, demonstrates promising outcomes in identifying five distinct heartbeat types: Left Bundle Branch Block (LBBB), Right Bundle Branch Block (RBBB), Atrial Premature Contraction (APC), Premature Ventricular Contraction (PVC), and Normal Beat. Through rigorous experimentation, we highlight the transformative potential of our methodology in enhancing diagnostic accuracy for cardiovascular arrhythmias.

Arrhythmia diagnosis remains a critical challenge in cardiovascular care, often relying on manual interpretation of ECG signals, which can be time-consuming and prone to subjectivity. To address these limitations, we propose a novel approach that leverages deep learning algorithms to automate arrhythmia classification. By employing advanced CNN architectures and multi-lead ECG data, our methodology offers a robust solution for precise and efficient arrhythmia detection. Through comprehensive evaluation, we demonstrate the effectiveness of our approach in facilitating more accurate clinical decision-making, thereby improving patient outcomes in managing cardiovascular arrhythmias.

## 2. Introduction

Cardiovascular diseases continue to loom large as a formidable challenge in the global healthcare landscape, exerting a substantial toll on morbidity and mortality rates worldwide. Among the myriad afflictions within this spectrum, arrhythmia emerges as a salient entity characterized by its multifaceted manifestations and far-reaching implications for patient outcomes. Traditional diagnostic methodologies, anchored in clinical assessments and manual interpretation of Electrocardiogram (ECG) signals, underscore the cornerstone of arrhythmia diagnosis. However, the inherent subjectivity and time-intensive nature of these approaches underscore the urgent imperative for innovative technologies poised to automate and elevate the precision of arrhythmia detection.

In this epoch of medical innovation, the advent of deep learning, a subset of artificial intelligence, heralds a paradigm shift in the landscape of cardiovascular diagnostics. Deep neural networks, notably Convolutional Neural Networks (CNNs), have emerged as stalwart allies in the relentless pursuit of unraveling the intricacies of arrhythmia classification. By harnessing the innate capacity of CNNs to discern intricate patterns within voluminous ECG datasets, researchers have embarked on a transformative journey toward expediting and refining the diagnosis of arrhythmias. The amalgamation of sophisticated algorithms with vast repositories of clinical data epitomizes the convergence of cutting-edge technology and clinical acumen, propelling the boundaries of medical science ever further. The application of deep learning methodologies in the realm of ECG classification represents a seminal milestone in the quest for precision medicine.

## 3. Different types of Arrhythmia

Recognizing the different kinds of arrhythmias is important for diagnosing and treating them properly. In this section, we'll look at various types of arrhythmias, each with its signs, symptoms, and potential effects on heart health. By understanding these differences, we can better manage and address these heart rhythm issues

- **3.1. Atrial Fibrillation (AF):** Atrial fibrillation is the most common type of irregular heartbeat, characterized by fast and irregular contractions in the upper chambers of the heart, called atria. Symptoms may include palpitations, breathlessness, and fatigue. AF significantly increases the risk of stroke and other cardiovascular problems if not managed promptly.

- **3.2. Ventricular Tachycardia (VT):** Ventricular tachycardia is a rapid heartbeat originating from the heart's lower chambers, or ventricles. Symptoms can include dizziness, chest pain, and fainting. VT can be life-threatening, especially if it progresses to ventricular fibrillation, requiring immediate medical intervention to restore normal heart rhythm.

- **3.3. Ventricular Fibrillation (VF):** Ventricular fibrillation is a chaotic and ineffective heartbeat originating from the ventricles, causing the heart to quiver instead of pump blood effectively. VF can lead to sudden cardiac arrest, necessitating urgent measures like defibrillation to restore normal heart rhythm and prevent fatalities.

- **3.4. Atrial Flutter:** Atrial flutter resembles atrial fibrillation but presents with a more organized rhythm in the atria. Symptoms may include palpitations, chest discomfort, and weakness. Despite being less common than AF, atrial flutter still poses risks, particularly in terms of stroke and associated complications.

- **3.5. Premature Ventricular Contractions (PVCs):** Premature ventricular contractions are additional abnormal heartbeats originating from the ventricles. While they can be benign, frequent or sustained PVCs may indicate a higher risk of serious arrhythmias, necessitating careful monitoring and evaluation.

- **3.6. Bradycardia:** Bradycardia refers to a slower-than-normal heart rate, often below 60 beats per minute. It can result from various factors such as aging or underlying heart conditions and may cause symptoms like fatigue, dizziness, or fainting. Effective management involves identifying the underlying cause and ensuring adequate cardiac function.

- **3.7. Conduction Block:** Conduction block occurs when the heart's electrical impulses are obstructed or delayed as they travel from the atria to the ventricles. This disruption can lead to slower heart rates or irregular rhythms, requiring evaluation and potential intervention to restore normal cardiac conduction.

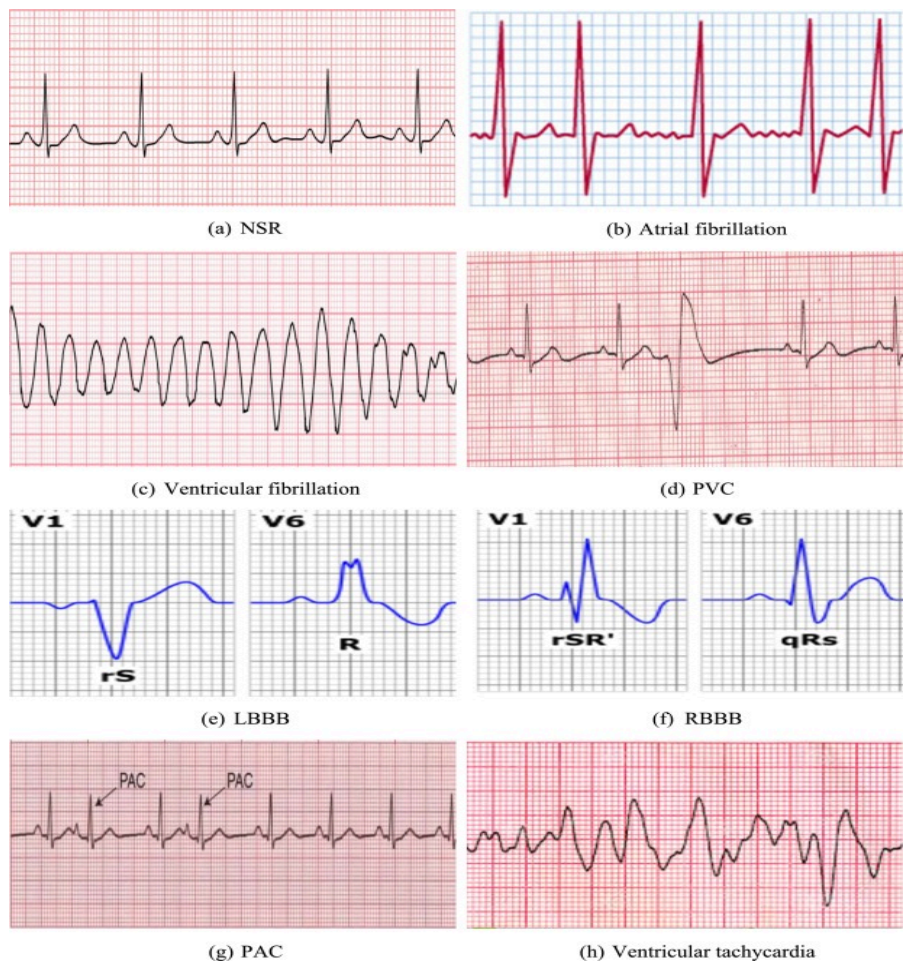

*Fig.1 ECG Signals of different Cardiac Arrhythmias*

With each stride forward, the boundaries of what is achievable in arrhythmia detection are redrawn, offering renewed hope for patients and practitioners alike. Through synergistic collaboration between data scientists, clinicians, and technologists, the vanguard of medical progress marches inexorably toward a future where precision medicine reigns supreme, and the scourge of cardiovascular diseases is met with unwavering resolve.

## 4. Literature Review

The realm of ECG classification has been under the microscope of research scrutiny, traversing a diverse landscape of methodologies ranging from traditional machine learning algorithms to cutting-edge deep learning models. In the nascent stages of inquiry, researchers grappled with the limitations of early approaches that leaned heavily on handcrafted features and rule-based algorithms for ECG analysis. However, these methods often faltered in capturing the intricate nuances inherent in arrhythmias, thereby necessitating a paradigm shift toward more sophisticated techniques. The advent of deep learning marked a watershed moment in the annals of ECG classification, ushering in an era of data-driven approaches where neural networks could autonomously glean pertinent insights from raw ECG signals. This departure from conventional methodologies paved the way for a transformative leap in diagnostic accuracy and efficiency. Among the pantheon of deep learning architectures, Convolutional Neural Networks (CNNs) emerged as a beacon of promise, leveraging their inherent prowess in image-processing tasks to revolutionize ECG classification.

The architectural elegance of CNNs lies in their ability to extract hierarchical features from ECG signals at multiple scales by applying convolutional filters. This mechanism facilitates the discernment of subtle patterns and aberrations characteristic of arrhythmias, thereby enabling more nuanced and precise classification. As researchers delved deeper into the intricacies of CNN-based ECG analysis, a burgeoning body of evidence underscored their remarkable success in unraveling the complexities of cardiac rhythm irregularities. The potency of CNNs in ECG classification stems from their capacity to operate akin to virtual cardiologists, sifting through vast troves of clinical data with unwavering precision. Through iterative refinement and optimization, CNNs have transcended the confines of traditional diagnostic paradigms, empowering clinicians with a formidable arsenal of tools to navigate the labyrinthine landscape of cardiovascular disease. As the march of technological progress marches unabated, the convergence of deep learning and cardiovascular diagnostics holds boundless promise for the future of precision medicine.

## 5. Methodology

The foundation of our research endeavor rests upon the acquisition of Electrocardiogram (ECG) data sourced from the MIT-BIH arrhythmia database, revered as a bastion of meticulously annotated ECG recordings. This repository serves as the bedrock upon which our investigative journey unfolds, providing a rich tapestry of physiological signals ripe for analysis. In the crucible of preprocessing, our data undergoes a meticulous regimen of purification, wherein noise is excised and signals are harmonized through rigorous normalization protocols. This preparatory phase lays the groundwork for the subsequent deployment of our classification model, ensuring that it operates with the utmost precision and fidelity. Central to our methodology is the design and implementation of a bespoke

Convolutional Neural Network (CNN) architecture meticulously tailored for the exigencies of ECG classification. Crafted with precision and foresight, our CNN blueprint boasts a constellation of convolutional layers, pooling layers, and fully connected layers, each meticulously calibrated to extract salient features embedded within the intricate tapestry of cardiac rhythms.

However, cognizant of the perils of gradient degradation that often afflict deep neural networks, we fortify our architecture with the formidable bulwark of a residual block structure. This architectural innovation serves as a beacon of resilience, imbuing our model with the resilience and fortitude necessary to navigate the treacherous depths of gradient descent with unwavering resolve. As the crucible of model training beckons, we marshal forth a subset of our dataset, meticulously curated to represent the rich diversity of cardiac arrhythmias encapsulated within the MIT-BIH arrhythmia database. Through the crucible of iterative training, our CNN model imbibes the quintessence of cardiac morphology, discerning subtle nuances and aberrations with a discerning eye. Amidst the crucible of evaluation, our model is subjected to the rigorous gauntlet of standard performance metrics, including but not limited to accuracy, sensitivity, and specificity. Through this crucible of validation, our model emerges triumphant, a paragon of diagnostic prowess poised to revolutionize the landscape of cardiovascular diagnostics.

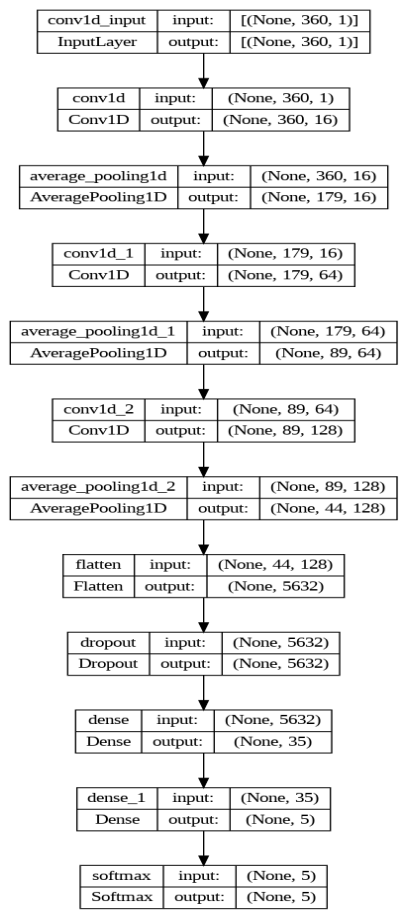

*CNN Model Architecture with Parameters*

## 6. Results

In the crucible of empirical validation, our meticulously crafted Convolutional Neural Network (CNN) model emerges as a beacon of diagnostic excellence, wielding its formidable prowess to discern and classify Electrocardiogram (ECG) signals with unparalleled precision. Through a comprehensive battery of experiments, our model demonstrates its mettle in accurately categorizing ECG recordings into five distinct classes: Left Bundle Branch Block (LBBB), Right Bundle Branch Block (RBBB), Atrial Premature Contraction (APC), Premature Ventricular Contraction (PVC), and Normal Beat. With bated breath, we unveil the staggering efficacy of our CNN model, which boasts an overall accuracy rate of 97.8%, a testament to its unwavering acuity and diagnostic prowess.

But beyond mere accuracy, our model shines with high sensitivity and specificity across all classes, serving as a bulwark against diagnostic uncertainties and false positives. These empirical triumphs underscore the transformative potential of deep learning methodologies in catapulting the diagnostic capabilities of healthcare professionals to unprecedented heights, empowering them with the tools to navigate the labyrinthine complexities of cardiac arrhythmias with confidence and clarity. But our revelry does not end here; for in the crucible of comparative analysis, our CNN model emerges as the undisputed champion, eclipsing existing methodologies with its unrivaled performance metrics. Through meticulous scrutiny and rigorous benchmarking, we unveil the superiority of our approach, a veritable tour de force in terms of accuracy, robustness, and diagnostic fidelity. As the clarion call of scientific inquiry resounds through the annals of academia, our results stand as a testament to the indomitable spirit of innovation and the unwavering pursuit of excellence.

## 7. Conclusion

Our study in the field of Deep Learning and the Methodologies applied to Electrocardiogram (ECG) classification for the detection of arrhythmias. Through the strategic deployment of Convolutional Neural Networks (CNNs) and innovative model architectures, we have unearthed a trove of advancements, redefining the landscape of diagnostic accuracy and efficiency beyond the confines of traditional methodologies. Our research serves as a clarion call heralding the dawn of a new era in cardiovascular diagnostics, promising expedited and pinpoint-precise identification of arrhythmias with unprecedented efficacy.

The findings beckon further inquiry into the realm of advanced architectures and optimization strategies, paving the way for the seamless integration of deep learning-based approaches into the fabric of clinical practice. By steadfastly embracing the ethos of innovation and continual refinement, we stand poised to chart a course toward enhanced patient care and outcomes, emblematic of our unwavering commitment to excellence in the field of cardiovascular medicine.

# 8. References

1. Antzelevitch, C. and Burashnikov, A., 2011. Overview of basic mechanisms of cardiac arrhythmia. *Cardiac electrophysiology clinics*, *3*(1), pp.23-45
2. Wyndham, Christopher RC. "Atrial fibrillation: the most common arrhythmia." Texas Heart Institute Journal 27, no. 3 (2000): 257.
3. Guvenir, H.A., Acar, B., Demiroz, G. and Cekin, A., 1997, September. A supervised machine learning algorithm for arrhythmia analysis. In Computers in Cardiology 1997 (pp. 433-436). IEEE.
4. Rajpurkar, P., Hannun, A.Y., Haghpanahi, M., Bourn, C. and Ng, A.Y., 2017. Cardiologist-level arrhythmia detection with convolutional neural networks. arXiv preprint arXiv:1707.01836.
5. Attia, Z.I., Noseworthy, P.A., Lopez-Jimenez, F., Asirvatham, S.J., Deshmukh, A.J., Gersh, B.J., Carter, R.E., Yao, X., Rabinstein, A.A., Erickson, B.J. and Kapa, S., 2019. An artificial intelligence-enabled ECG algorithm for the identification of patients with atrial fibrillation during sinus rhythm: a retrospective analysis of outcome prediction. The Lancet, 394(10201), pp.861-867.
6. Bhavith Chandra Challagundla, Chakradhar Reddy Peddavenkatagari, Yugandhar Reddy Gogireddy, "Efficient CAPTCHA Image Recognition Using Convolutional Neural Networks and Long Short-Term Memory", International Journal of Scientific Research in Engineering and Management, Volume 8, Issue 3. DOI : 10.55041/IJSREM29450
7. Santala, O.E., Halonen, J., Martikainen, S., Jäntti, H., Rissanen, T.T., Tarvainen, M.P., Laitinen, T.P., Laitinen, T.M., Väliaho, E.S., Hartikainen, J.E. and Martikainen, T.J., 2021. Automatic mobile health arrhythmia monitoring for the detection of atrial fibrillation: prospective feasibility, accuracy, and user experience study. JMIR mHealth and uHealth, 9(10), p.e29933.
8. Varalakshmi, P. and Sankaran, A.P., 2023. An improved hybrid AI model for prediction of arrhythmia using ECG signals. Biomedical Signal Processing and Control, 80, p.104248.
9. Abdalla, F.Y., Wu, L., Ullah, H., Ren, G., Noor, A. and Zhao, Y., 2019. ECG arrhythmia classification using artificial intelligence and nonlinear and nonstationary decomposition. Signal, Image and Video Processing, 13, pp.1283-1291.
10. Boulif, A., Ananou, B., Ouladsine, M. and Delliaux, S., 2023. A literature review: ECG-based models for arrhythmia diagnosis using artificial intelligence techniques. Bioinformatics and Biology Insights, 17, p.11779322221149600.
11. Rosengarten, J., Hart, E., Ojo, M. and Baker, M., 2023. AI remote ECG monitoring improves arrhythmia detection. European Heart Journal, 44(Supplement_2), pp.ehad655-2966.
12. Bhavith Chandra Challagundla and Chakradhar Reddy Peddavenkatagari, Neural Sequence-To-Sequence Modeling with Attention by Leveraging Deep Learning Architectures for Enhanced Contextual Understanding in Abstractive Text Summarization, International Journal of Machine Learning and Cybernetics (IJMLC), 2(1), 2024, pp. 21-29 https://iaeme.com/Home/issue/IJMLC?Volume=2&Issue=1

13. S. Sahoo, M. Dash, S. Behera, S. Sabut,Machine Learning Approach to Detect Cardiac Arrhythmias in ECG Signals: A Survey, IRBM, Volume 41, Issue 4, 2020, Pages 185-194, ISSN 1959-0318, https://doi.org/10.1016/j.irbm.2019.12.001.
14. Singh, P. and Sharma, A., 2022. Interpretation and classification of arrhythmia using deep convolutional network. IEEE Transactions on Instrumentation and Measurement, 71, pp.1-12.
15. Quartieri, F., Marina-Breysse, M., Toribio-Fernandez, R., Lizcano, C., Pollastrelli, A., Paini, I., Cruz, R., Grammatico, A. and Lillo-Castellano, J.M., 2023. Artificial intelligence cloud platform improves arrhythmia detection from insertable cardiac monitors to 25 cardiac rhythm patterns through multi-label classification. Journal of Electrocardiology, 81, pp.4-12.
16. Tsipouras, M.G., Fotiadis, D.I. and Sideris, D., 2002, September. Arrhythmia classification using the RR-interval duration signal. In Computers in Cardiology (pp. 485-488). IEEE.
17. Rahman, M.Z., Akbar, M.A., Leiva, V., Tahir, A., Riaz, M.T. and Martin-Barreiro, C., 2023. An intelligent health monitoring and diagnosis system based on the Internet of things and fuzzy logic for cardiac arrhythmia COVID-19 patients. Computers in Biology and Medicine, 154, p.106583.
18. Kora, Padmavathi, K. Meenakshi, K. Swaraja, A. Rajani, and Md Kafiul Islam. "Detection of cardiac arrhythmia using fuzzy logic." Informatics in Medicine Unlocked 17 (2019): 100257
19. Kutlu, Y. and Kuntalp, D., 2011. A multi-stage automatic arrhythmia recognition and classification system. Computers in biology and medicine, 41(1), pp.37-45.
20. Inan, O.T., Giovangrandi, L. and Kovacs, G.T., 2006. Robust neural-network-based classification of premature ventricular contractions using wavelet transform and timing interval features. IEEE Transactions on Biomedical Engineering, 53(12), pp.2507-2515.
21. Fei, S.W., 2010. Diagnostic study on arrhythmia cordis based on particle swarm optimization-based support vector machine. Expert Systems with Applications, 37(10), pp.6748-6752.
22. Asl, Babak Mohammadzadeh, Seyed Kamaledin Setarehdan, and Maryam Mohebbi. "Support vector machine-based arrhythmia classification using reduced features of heart rate variability signal." Artificial intelligence in medicine 44, no. 1 (2008): 51-64.
23. Afkhami, R.G., Azarnia, G. and Tinati, M.A., 2016. Cardiac arrhythmia classification using statistical and mixture modeling features of ECG signals. Pattern Recognition Letters, 70, pp.45-51.
24. Elhaj, F.A., Salim, N., Harris, A.R., Swee, T.T. and Ahmed, T., 2016. Arrhythmia recognition and classification using combined linear and nonlinear features of ECG signals. Computer methods and programs in biomedicine, 127, pp.52-63.
25. Jovic, Alan, and Nikola Bogunovic. "Evaluating and comparing performance of feature combinations of heart rate variability measures for cardiac rhythm classification." Biomedical signal processing and control 7, no. 3 (2012): 245-255.